\documentclass[aps,pre,twocolumn,floats,showpacs]{revtex4}
\usepackage{epsfig}
\usepackage{bm}
\usepackage{latexsym}

\begin{document}
\newcommand{\hide}[1]{}
\newcommand{\tbox}[1]{\mbox{\tiny #1}}
\newcommand{\half}{\mbox{\small $\frac{1}{2}$}}
\newcommand{\sinc}{\mbox{sinc}}
\newcommand{\const}{\mbox{const}}
\newcommand{\trc}{\mbox{trace}}
\newcommand{\intt}{\int\!\!\!\!\int }
\newcommand{\ointt}{\int\!\!\!\!\int\!\!\!\!\!\circ\ }
\newcommand{\eexp}{\mbox{e}^}
\newcommand{\bra}{\left\langle}
\newcommand{\ket}{\right\rangle}
\newcommand{\EPS} {\mbox{\LARGE $\epsilon$}}
\newcommand{\ar}{\mathsf r}
\newcommand{\im}{\mbox{Im}}
\newcommand{\re}{\mbox{Re}}
\newcommand{\bmsf}[1]{\bm{\mathsf{#1}}}
\newcommand{\mpg}[2][1.0\hsize]{\begin{minipage}[b]{#1}{#2}\end{minipage}}


\title{Parametric invariant Random Matrix Model and the emergence
of multifractality}

\author{
J. A. M\'endez-Berm\'udez$^{1,2,3}$, Tsampikos Kottos$^{1,4}$, Doron Cohen$^{2}$
}

\affiliation{
$^1$Max-Planck-Institut f\"ur Dynamik und Selbstorganisation, Bunsenstra\ss e 10,
D-37073 G\"ottingen, Germany \\
$^2$Department of Physics, Ben-Gurion University,
Beer-Sheva 84105, Israel \\
$^3$Instituto de F\'{\i}sica, Universidad Aut\'onoma de Puebla, Apartado
Postal J-48, Puebla 72570, M\'exico \\
$^4$Department of Physics, Wesleyan University, Middletown, Connecticut
06459-0155, USA
}

\begin{abstract}

We propose a random matrix modeling
for the parametric evolution of eigenstates.
The model is inspired by a large class
of quantized chaotic systems. Its unique
feature is having parametric invariance
while still possessing the non-perturbative breakdown
that has been discussed by Wigner 50~years ago.
Of particular interest is the emergence
of an additional crossover to multifractality.

\end{abstract}
\maketitle


\section{Introduction}

The analysis of structural changes of eigenstates as
a parameter $x$ is varied, has sparked a great deal
of research activity for many years
\cite{W55,BCH,MKC05,MLI04,B03,VLG02,CH2000,CK01,HKG06}.
Of particular importance are quantized {\em chaotic}
or complex systems where the change of $x$ may represent
the effect of some externally controlled field (like
electric field, gate voltage or magnetic flux). Thus,
these studies are relevant for diverse areas of physics
ranging from nuclear \cite{HZB95,W55} and atomic physics
\cite{FGGP99} to quantum chaos \cite{CK01,MLI04,B03}
and mesoscopics \cite{VLG02}.

In all these studies, Random Matrix Theory (RMT) played
a dominant role as a reference theory that describes the
{\it universal} properties of the eigenstates of complex
systems. RMT was introduced 50 years ago by Wigner as a
tool to describe the statistical properties of the
eigenvalues of complex nuclei.
Until recently, the matrices in the frame of RMT were
assumed to be homogeneous, i.e., all matrix elements were
set to have identical statistical properties. Under this
simplification random matrices are rotationally invariant,
a property that simplifies their theoretical analysis. In
physical applications this implies that interactions are
assumed to be so strong and complex that no other parameters,
apart from the symmetry of the Hamiltonian matrix, are
relevant. As a result, such random matrices can be associated
to the extreme case of maximal chaos, which is known to appear
in various physical systems such as heavy nuclei, atoms,
metallic clusters, etc. Moreover, one can treat full random
matrices as a typical model when describing local statistical
properties of spectra and eigenstates in some range of the
energy spectrum, typically, in the semiclassical region.

On the other hand, the conventional RMT can not describe
important phenomena such as localization of eigenstates,
neither can be directly applied to obtain spectra of
realistic models. For this reason, much attention has been
recently paid to the so-called Wigner Band Random Matrix
(WBRM) model which is characterized by the free parameter
$b$, that represents the effective bandwidth of a
Hamiltonian matrix.
Among the important applications of the WBRM model we
mention the study of localization in quasi-one-dimensional
disordered systems. Also the WBRM model has been applied
to the analysis of either {\em chaotic} or complex
conservative quantum systems that are present
in nuclear physics as well as in atomic and molecular physics.

Despite its success, the standard WBRM model 
has {\em severe limitations in modeling realistic 
systems}. We explain these limitations in section II, 
and further motivate the introduction of  
a new RMT ensemble to which we refer as 
the Winger Lorentzian Random Matrix (WLRM) model.
In Sec. III, the WLRM model is shown 
to have a ``parametric~invariance" property 
which is characteristic of any realistic system
but is missing in the standard WBRM model.
The analysis of the local density of states 
of the WLRM model is done in Sec. IV. 
The multifractal properties of the eigenstates
of the WLRM ensemble are analyzed in Sec.V. 
Our conclusions are summarized in Sec. VI.

\section{RMT modeling}

The pioneering work in this field has been done
by Wigner \cite{W55}, who has motivated the
studies of RMT models of the type
\begin{equation}
{\cal H}=\bm{E}+x\bm{B} \ .
\label{ham}
\end{equation}
Both $\bm{E}$ and $\bm{B}$ are real symmetric matrices
of size $N\times N$. The elements of the diagonal
matrix $\bm{E}$ are the ordered energies $\{E_n\}$,
with mean level spacing $\Delta$,
while $\bm{B}$ is a banded {\em random} matrix which
is characterized by a band profile $C(r)$.
Namely, the entries of $\bm{B}$ are random numbers
that are drawn from a normal distribution with zero
mean and variance given by
\begin{equation}
\langle  |B_{nm}|^2 \rangle = C(n-m) \ .
\end{equation}
For the study of {\em spectral statistics}
of the energy levels it turns out that full matrices ($C(r)=1$),
say of the Gaussian Orthogonal ensemble (GOE),
are enough in order to capture the universality
which is found in quantized chaotic system.
But for the study of {\em parametric evolution}
of the eigenstates it is essential to take into account
the band profile, which is dictated
by semiclassical considerations.
Namely, $C(r)$ is merely the scaled version
of a classical power spectrum $\tilde{C}(\omega)$
which is obtained via a Fourier transform
of the classical correlation function of the
generalized force ${\cal F}(t) = -\partial {\cal H}/\partial x$.

\subsection{The standard WBRM model}

The standard WBRM model assumes a rectangular band profile:
\begin{equation}
C(r)= \left\{ \matrix{
1  & \quad \mbox{$r\le b$} \cr
0  & \quad \mbox{$r> b$}
} \right. \ .
\label{WBRM}
\end{equation}
For this model Wigner has found that the eigenstates undergo
a transition from a {\em perturbative} Lorentzian-type line
shape to a {\em non-perturbative} semicircle line-shape.
It should be clear that the Wigner Lorentzian can be regarded
as the outcome of perturbation theory to infinite order,
while the semicircle line-shape is beyond any order
of perturbation theory.

The {\em existence of a transition} from a perturbative
to a non-perturbative line shape is a generic feature
of {\em any} realistic (quantized) Hamiltonian.
In the latter case the {\em semi}-circle line-shape
is replaced by a {\em semi}-classical line shape.

\subsection{The modified WBRM model}

The WBRM model suffers from a serious drawback.
Unlike generic canonically quantized Hamiltonians,
the statistical properties of its Hamiltonian
matrix are not invariant under $x \rightarrow x + \const$.
In fact there exists two wise modified versions
of the WBRM model \cite{Alhassid,Wilkinson}
which are manifestly $x$~invariant by construction.
For example we cite one of them:
\begin{equation}
\mathcal{H} = \bm{E}+\cos(x)\bm{B}_1 + \sin(x)\bm{B}_2
\label{modWLRM}
\end{equation}
Here $\bm{B}_1$ and $\bm{B}_2$ are uncorrelated
banded matrices. It is quite easy to be convinced
that this model is $x$~invariant. One simply has
to set $x \rightarrow x + \const$, to expand the $\sin()$
and the $\cos()$, to define $\bm{B}_1'$ and $\bm{B}_2'$,
and to observe that they are uncorrelated
with the same band profile as $\bm{B}_1$ and $\bm{B}_2$.

However there is a ``price" for using such modified model.
It is not difficult to prove that the parametric
nature of this model is essentially perturbative:
The associated local density of states does
{\em not} exhibit the non-perturbative crossover that
has been highlighted in the previous subsection!

\subsection{The Winger Lorentzian Random Matrix model}

In the present paper we introduce a
new RMT ensemble to which we refer as
the Winger Lorentzian Random Matrix (WLRM) model.
The Hamiltonian is assumed to have the standard
form of Eq.(\ref{ham}), and it is characterized
by the band profile
\begin{equation}
C(r) =
\frac{1}{1+\left( r/b \right)^2} \, .
\label{PBRM}
\end{equation}
There are several good reasons that motivate
the introduction and the study of this model,
which we are going to clarify:
\begin{itemize}
\item[{\bf (1)}]
There is a major class of quantized
chaotic systems that can be described using
this model.
\item[{\bf (2)}] Unlike the standard WBRM model it has
the desired $x$~invariance property that
characterizes quantized models.
\item[{\bf (3)}] Unlike the common $x$~invariant
version of the WBRM model it exhibits the transition
to a non-perturbative line shape.
\item[{\bf (4)}] The emergence of multifractality,
which is absent in the WBRM model,
is a fascinating issue by itself.
\end{itemize}

Let us expand on the first point.  We recall that
the flat band profile of the standard WBRM model
is motivated by the realization that many observables
(say ${\cal F}(t)$) of chaotic systems exhibit ``white" power 
spectrum: $\tilde{C}_{{\cal FF}}(\omega)\sim \mbox{const}$. 
However, in many cases it is $G(t)=\dot{\cal F}$ that 
has the ``white" power spectrum \cite{BCH,MKC05}.
In the latter case the relation 
$\tilde{C}_{{\cal FF}}(\omega)=\tilde{C}_{GG}(\omega)/\omega^2$ 
implies Lorentzian tails.

\hide{
A major example is the Aharonov-Bohm cylindrical
billiard \cite{MKC05}. The motion on the two dimensional
surface of this system is chaotic due to the
collisions of the particle with a deformed or possibly rough
upper boundary.  The control parameter $x$
is the magnetic flux $\Phi$ through the cylinder.
The corresponding Hamiltonian has the form (\ref{ham}),
where $\bf{B}_{nm}$ are the matrix elements of the
current operator.  The band profile of $B$ is of
the Lorentzian type, which reflects the velocity-velocity
correlation function: Namely, the associate power spectrum
$\tilde{C}_{vv}(\omega)$ possess $1/omega^2$ tails.
It follows because the velocity ($v$) is the time derivative
of a random like force which is exerted on the particle
due to collisions with the walls.
}

\section{Parametric invariance}

\begin{figure}[t]
\begin{center}
    \epsfxsize=8cm
    \leavevmode
    \epsffile{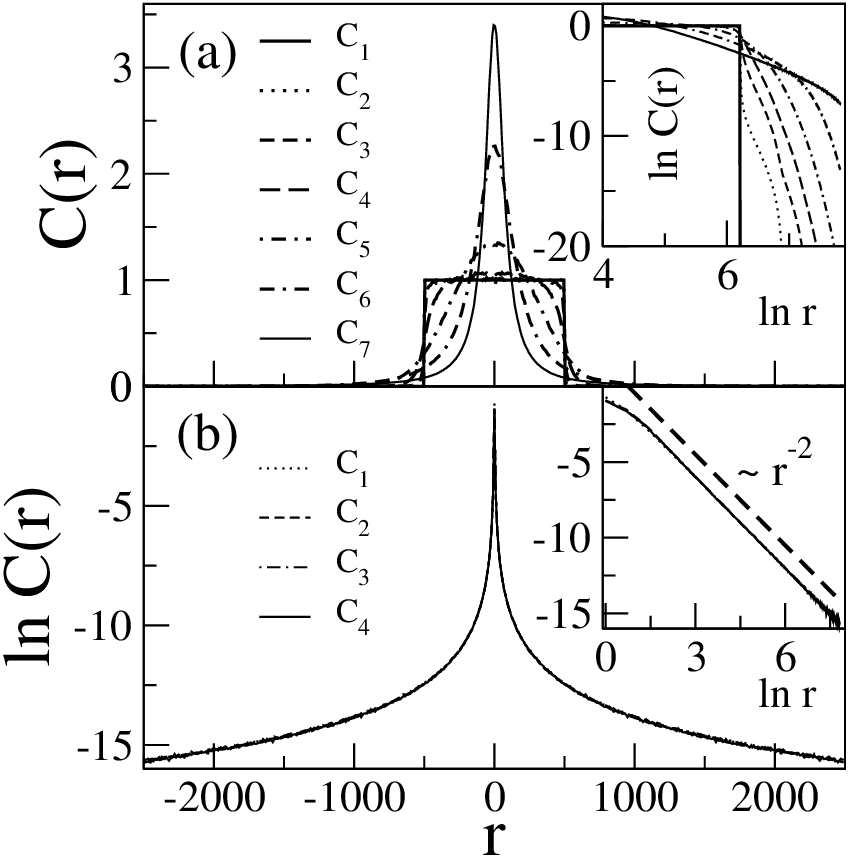}
\caption{(a) The band profile $C(r)$ of the standard Wigner model
using $b=500$ and $x^*=0$, 0.1, 1, 3, 7, 15, 31, and 63.
(b) The band profile $C(r)$ of the Lorentzian Wigner model
with $b=1$ and $x^*=2.5$, 27, 277, and 2777.
The dashed line in the inset of (b) with decay $\sim r^{-2}$
is plotted to guide the eye.}
\label{fig:fig1}
\end{center}
\end{figure}

\begin{figure}[t]
\begin{center}
    \epsfxsize=8cm
    \leavevmode
    \epsffile{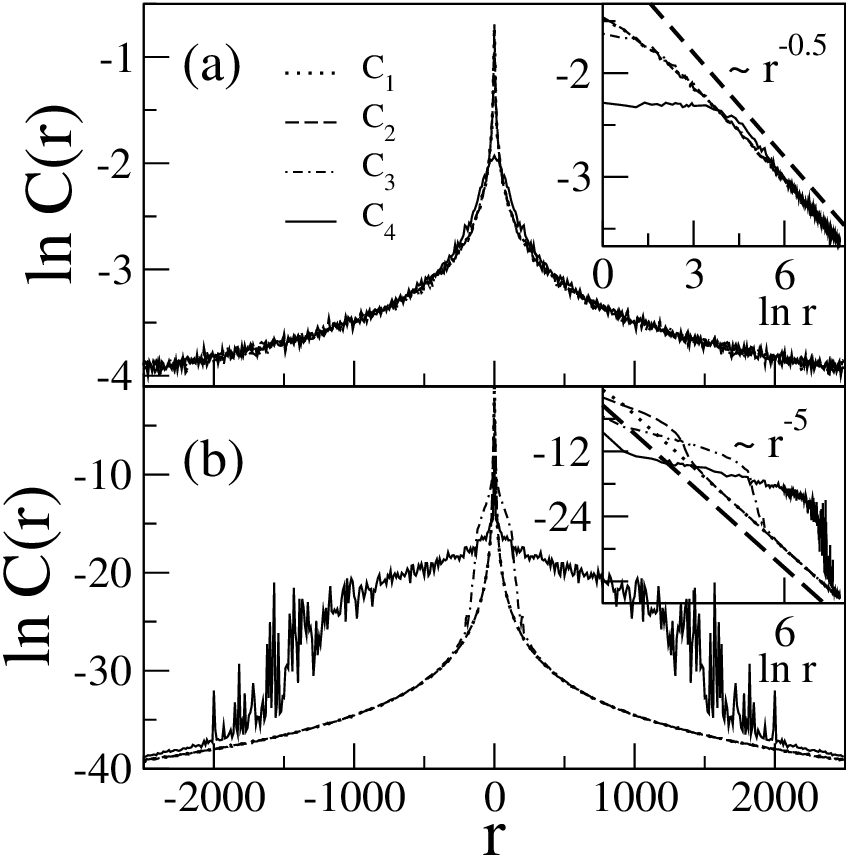}
\caption{The band profiles for the power law matrix
$C(r) = [1+(r/b)^\mu]^{-1}$ with (a) $\mu=0.5$ and (b) 5.
$b=1$. In (a) $x^*=2.5$, 27, 277, and 2777.
In (b) $x^*=0.13$, 1.4, 14, and 144.
The dashed line in the inset of (a) [(b)] with decay
$\sim r^{-0.5}$ [$\sim r^{-5}$] is plotted to guide the eye.}
\label{fig:fig2}
\end{center}
\end{figure}

An important feature of a generic canonically quantized
Hamiltonian $\mathcal{H}(Q,P;x)$ is its parametric $x$~invariance.
Given $x^*$ we can represent the Hamiltonian
by a matrix
\begin{eqnarray}
\mathcal{H} \ \ = \ \  \bm{E} + \delta x\bm{B}
\end{eqnarray}
where $\delta x = x-x^*$.
If we take two different values of $x^*$
we get two different $\bm{B}$ matrices.
But if the two values of $x^*$ belong
to the same {\em classically small window},
then (by definition) the band profile $C(r)$
comes out the same. Still from a quantum mechanical
point of view a classically small range
of $x$~values is typically regarded as {\em huge}.
This means that in general, quantum perturbation theory
cannot be used in order to describe the parametric
evolution within this range.

In Fig.~\ref{fig:fig1} we present $C(r)$
for the standard WBRM model and for the WLRM model.
We see that the profile of the perturbation matrix $\bm{B}$ of
the WBRM model is deformed as $x$ is increased, while that of
the WLRM model remains the same. We have found out that
this $x$~invariance does not hold for other (non-Lorentzian)
power law profiles. As examples, in Fig.~\ref{fig:fig2}
we present the profiles of $\bm{B}$ for increasing $x$ in the
case of the power-law profile $C(r) = [1+(r/b)^\mu]^{-1}$ with
$\mu=0.5$ and 5. We conclude that the $x$~invariance is a
unique property of the WLRM ensemble.

\section{LDOS analysis}

The local density of states (LDOS) is the major tool
for the characterization of the parametric evolution
of the eigenstates. The overlap of the eigenstates
$|n(x)\rangle$ for a given value of $x$ with the
eigenstate $|m(0)\rangle$ of the $x=0$ Hamiltonian
is $P(n|m) = |\langle n(x)|m(0)\rangle|^2$.
This can be regarded as a distribution with respect
to $n$. By averaging over the reference level $m$
we get the line shape $P(n-m)$. Up to trivial scaling
this is the LDOS.

The considerations that are required in order to generalize
the calculation of the LDOS line shape for a general band
profile have been introduced in \cite{CH2000,CK01}. Here, we
apply such methodology in order to analyze the parametric
evolution of the LDOS for the WLRM model.

For $x=0$ the LDOS is trivially ${P(r)=\delta_{r,0}}$ due to
orthogonality. As $x$ increases, perturbative tails start to appear.
By employing standard first-order perturbation
theory we get $P_{\tbox{FOPT}} (r) \approx 1$ for $r=0$, while
\begin{equation}
P_{\tbox{FOPT}}(r) = {x^2 |\bm{B}_{nm}|^2 \over (E_n{-}E_m)^2}
= \frac{x^2}{\Delta^2} \frac{b^2}{(b^2+r^2)} \frac{1}{r^2}
\label{Pst}
\end{equation}
for $r\neq 0$. In the second equality we have 
substituted the matrix elements of $\bm{B}$
using Eq.~(\ref{PBRM}). The above expression applies 
for $x<x_c$ where $x_c$ is the perturbation
strength needed to mix neighboring levels only.

For $x>x_c$ Wigner had found, in the frame of the WBRM model,
that the LDOS line shape can
be calculated using perturbation theory to infinite order. 
In case of the WBRM model one
obtains a Lorentzian. Assuming the validity of infinite 
order perturbation theory we come
out with a Lorentzian-type approximation 
for the LDOS of the WLRM model:
\begin{eqnarray}
P_{\tbox{PRT}}(r) &=& {x^2 |\bm{B}_{nm}|^2 \over \Gamma^2 + (E_n{-}E_m)^2}
\nonumber\\
&=& \frac{x^2}{\Delta^2} \frac{b^2}{(b^2+r^2)} 
\frac{1}{[(\Gamma/\Delta)^2+r^2]} \ .
\label{Pw}
\end{eqnarray}
Eq.~(\ref{Pw}) is an approximation
because all orders of perturbation are treated
within a Markovian-like approach (by iterating
the first order result) and convergence of the expansion is pre-assumed.
The energy scale $\Gamma$ defines the region where a non-perturbative mixing of levels occurs.
Inside this region the perturbative profile $P_{\tbox{PRT}}(r)$ does not describe the actual LDOS
lineshape. $\Gamma$ is determined by imposing normalization of $P_{\tbox{PRT}}(r)$.
For the WBRM model it was found \cite{W55}
that $\Gamma= x^2 C(1)/\Delta$. For the WLRM model we get
\begin{equation}
\Gamma = \frac{b\Delta}{2} \left[ \sqrt{1+\frac{4\pi x^2}{b\Delta^2}} - 1\right].
\label{gamma}
\end{equation}

\begin{figure}[t]
\begin{center}
    \epsfxsize=8cm
    \leavevmode
    \epsffile{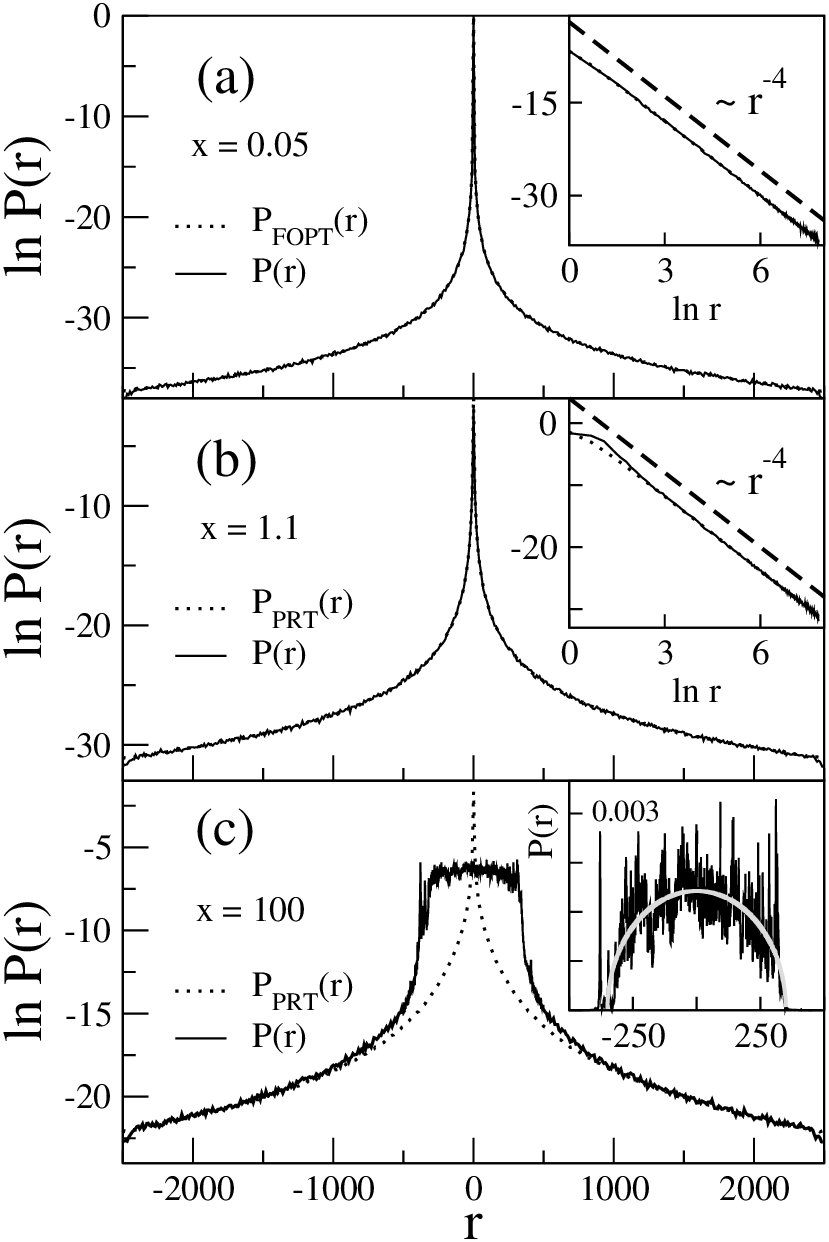}
\caption{The LDOS lineshape $P(r)$ in the (a) standard perturbative,
(b) extended perturbative, and (c) non-perturbative regimes for the
WLRM model with $b=1$, $\Delta=1$, and $N=5000$.
For this set of parameters $x_c\approx 0.8$ and $x_{\tbox{prt}}\approx 1.5$.
The first-order perturbation theory profile $P_{\tbox{FOPT}}(r)$ from
Eq.~(\ref{Pst}) is included in (a).
The Wigner-type lineshape $P_{\tbox{PRT}}(r)$ from
Eq.~(\ref{Pw}), which is expected to be valid in the perturbative
regime, is also included (no fitting parameters) in (b) and (c).
The dashed line in the insets of (a) and (b) with decay $\sim r^{-4}$
is plotted to guide the eye.
The gray line in the inset of (c) is a semicircle fitting to $P(r)$.}
\label{fig:fig3}
\end{center}
\end{figure}

Obviously, for $\Gamma\ll \Delta$ the (infinite-order)
LDOS profile $P_{\tbox{PRT}}(r)$ reduces
to the standard first-order perturbation
theory expression $P_{\tbox{FOPT}}(r)$. Therefore $x_c$
can be determined by the condition $\Gamma(x_c) \approx \Delta$,
leading to
\begin{equation}
x_c \approx \frac{\Delta}{\sqrt{\pi}} \sqrt{1+\frac{1}{b}} \, .
\label{xc}
\end{equation}

In Fig.~\ref{fig:fig3} we display our numerical results
for $P(r)$ for the WLRM model with
$b=1$, $\Delta=1$, $N=5000$, and various
perturbation strengths $x$.
We see that the agreement
with the perturbative expression (\ref{Pw})
persist up to some perturbation strength $x_{\tbox{prt}}$.
Above $x_{\tbox{prt}}$ the LDOS lineshape $P(r)$
becomes semicircle in complete analogy with the WBRM model scenario.

\begin{figure}[t]
\begin{center}
    \epsfxsize=8cm
    \leavevmode
    \epsffile{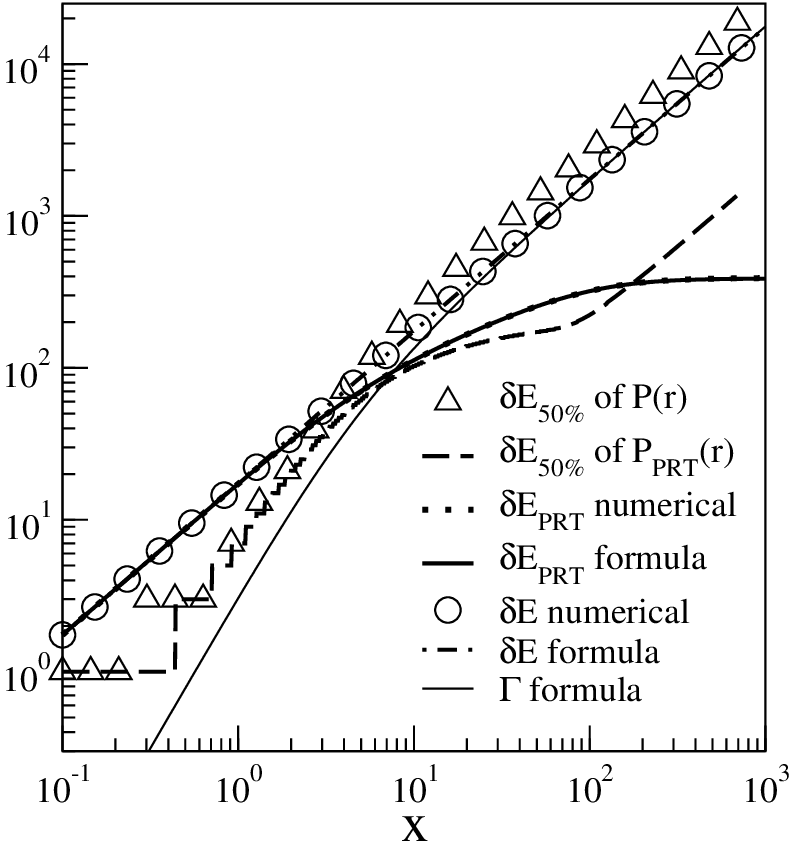}
\caption{$\delta E_{\tbox{PRT}}$ and $\delta E$ as a function of
$x$ for the WLRM model with $b=100$, $\Delta=1$, and $N=5000$.
The $50\%$ probability width of $P(r)$ ($\bigtriangleup$)
and of $P_{\tbox{PRT}}(r)$ (dashed line) are plotted to verify the
validity of
Eq.~(\ref{Pw}) for $x<x_{\tbox{prt}}$. $\delta E_{\tbox{PRT}}$
(dotted line) and $\delta E$ ($\circ$) were obtained from
$\delta E_{\tbox{PRT}}^2 = \Delta \sum_r r^2 P_{\tbox{PRT}}(r)$ and
$\delta E^2 = \Delta \sum_r r^2 P(r)$, respectively.
The numerically obtained $\delta E_{\tbox{PRT}}$ (dotted line)
is hardly visible in the plot because it lies on top of
$\delta E_{\tbox{PRT}}$ from Eq.~(\ref{dEW}).
We also show $\Gamma$ (thin full line) from Eq.~({\ref{gamma}}).
For this set
of parameters $x_c\approx 0.57$ and $x_{\tbox{prt}}\approx 5.35$.}
\label{fig:fig4}
\end{center}
\end{figure}

We want to find the value $x_{\tbox{prt}}$ up
to which the perturbative expression $P_{\tbox{PRT}}(r)$
describes reasonably good the LDOS lineshape.
To this end we compare the dispersion of $P_{\tbox{PRT}}(r)$,
\begin{equation}
\delta E_{\tbox{PRT}} = \Delta \times \sqrt{\sum_r r^2 P_{\tbox{PRT}}(r)} \ ,
\end{equation}
to the dispersion of the actual LDOS \cite{CK01}
\begin{equation}
\delta E = x\sum_{r\ne 0} C(r) \ .
\end{equation}
Expressions for $\delta E_{\tbox{PRT}}$ and $\delta E$
in case of our WLRM model can be obtained
by replacing the sums above by integrals:
\begin{equation}
\delta E_{\tbox{PRT}} \approx x\sqrt{\pi} \ b(b+\Gamma/\Delta)^{-1/2} \ ,
\label{dEW}
\end{equation}
\begin{equation}
\delta E \approx x\sqrt{2b} \left[ \pi/2 - \arctan(1/b) \right]^{1/2} .
\label{dE}
\end{equation}
Note that for small $x$, $\delta E_{\tbox{PRT}} = \delta E \approx x(\pi b)^{1/2}$.
$\delta E$ is a linear function of $x$ for all perturbation strengths
while for large enough $x$ $\delta E_{\tbox{PRT}}$ becomes sublinear:
$\delta E_{\tbox{PRT}} \propto x^{1/2}$. See Fig.~\ref{fig:fig4}.
The border $x_{\tbox{prt}}$ is identified as the perturbation strength for
which $\delta E_{\tbox{PRT}} (x_{\tbox{prt}}) \approx \gamma \delta E(x_{\tbox{prt}})$,
where $\gamma<1$:
\begin{equation}
x_{\tbox{prt}} \approx \Delta\sqrt{b}
\frac{\sqrt{\pi-2\gamma^2[\pi/2 - \arctan(1/b)]}}{2\gamma^2[\pi/2 - \arctan(1/b)]} \, .
\label{xprt}
\end{equation}
We typically use $\gamma = 0.8$.

The validity of Eqs.~(\ref{Pw}), (\ref{dEW}), and
(\ref{dE}) is confirmed in Fig.~\ref{fig:fig4} for $b=100$,
$\Delta=1$, and $N=5000$. There, we observe excellent
agreement between the expressions for $\delta E_{\tbox{PRT}}$
and $\delta E$ and the corresponding numerics.
To verify the validity of Eq.~(\ref{Pw}) we compare the
$50\%$ probability width (defined as the energy width of the
central $r$ region that contains $50\%$ of the probability)
of $P(r)$ to that of $P_{\tbox{PRT}}(r)$  finding reasonable
good agreement for $x<x_{\tbox{prt}}$, as expected.

Does the LDOS analysis capture all the features
of the eigenstates? The answer turns out to be negative.
In case of the WBRM model there is still another
regime which is not captured by the LDOS analysis.
Namely, for $x>x_{\infty}$ where $x_{\infty}=b^{3/2}x_c$
the eigenstates of the Hamiltonian become
exponentially localized. This is the well known
Anderson (strong) localization effect. For $x>x_{\infty}$
the Hamiltonian is essentially $\mathcal{H}=\bm{B}$
as if the diagonal $\bm{E}$ does not exist.
Do we have an analogous type of crossover in case
of the WLRM model?  The answer must be positive because
we know \cite{EM00} that in case of
the WLRM model, the Hamiltonian $\mathcal{H}=\bm{B}$
has multifractal (rather than localized) eigenstates.

\section{Multifractality}

The multifractal structure of the eigenstates is
commonly characterized by the fractal dimension $D_2$
which is associated with the scaling of the inverse
participation ratio. Given an eigenstate
of $\mathcal{H}$ which is represented by a column
vector $\Psi_n$ we define the participation ratio as
\begin{equation}
\mathcal{N}_2 = \left[ \sum_n |\Psi_n|^4 \right]^{-1} \ .
\end{equation}
The exponent $D_2$ is defined via the scaling relation
\begin{equation} \label{eDdef}
\overline{\mathcal{N}_2} \ \ \propto \ \ N^{D_2} \ .
\end{equation}
where ${\overline{\mathcal{N}_2} \equiv \exp(\langle \ln \mathcal{N}_2 \rangle)}$.
The measure $\overline{\mathcal{N}_2}$ constitutes an estimate
for the typical number of non-zero eigenfunction components
of the column vector. For a localized state it equals
a constant number and hence $D_2=0$, while for an
extended non-fractal state it is proportional
to the size of the matrix $N$ and hence $D_2=1$.
In general one finds $0<D_2<1$.
The fractal dimension $D_2$ manifests itself in a variety
of physical circumstances. As examples we mention the conductance
distribution in metals \cite{AKL91,BHMM01}, the statistical properties
of the spectrum \cite{BHMM01}, the anomalous spreading of a wave-packet,
the spatial dispersion of the diffusion coefficient \cite{HK99} and the
anomalous scaling of delay times \cite{MK05}.


In \cite{EM00} it was shown that the fluctuations
of $\mathcal{N}_2/\overline{\mathcal{N}_2}$ for $\mathcal{H}=\bm{B}$
are characterized by a universal probability distribution.
A theoretical estimation \cite{MFDQS96} gives
\begin{equation}
D_2 = \left\{ \matrix{
1-(\pi b)^{-1}  & \mbox{$b \gg 1$} \cr
2b & \mbox{$b \ll 1$}
} \right. \ . \nonumber
\end{equation}
We notice that $D_2=D(b)$ gives a global fit
for the fractal dimension, where we define
\begin{equation} \label{Db}
D(b) = \frac{1}{1+(2.34 b)^{-1}} \ .
\label{D2inf}
\end{equation}
Note that (\ref{D2inf}) is also in agreement
with the numerical found \cite{V02}
value ${D_2 \approx 0.7}$ for $b=1$.
For sake of later analysis we have found that
the associated proportionality
factor in Eq.~(\ref{eDdef}) is $\exp(-G(b))$ where
\begin{equation}
G(b) \approx \frac{1}{1+(1.23 b)^{-1}} \ .
\end{equation}


Thus, based on the above,
we may say that for $x=\infty$
we expect to have multifractal
eigenstates. We turn now to
discuss the more general case
of finite $x$. We want to see how
the multifractality emerges as
we increase $x$ form zero to infinity.

\begin{figure}[t]
\begin{center}
    \epsfxsize=8cm
    \leavevmode
    \epsffile{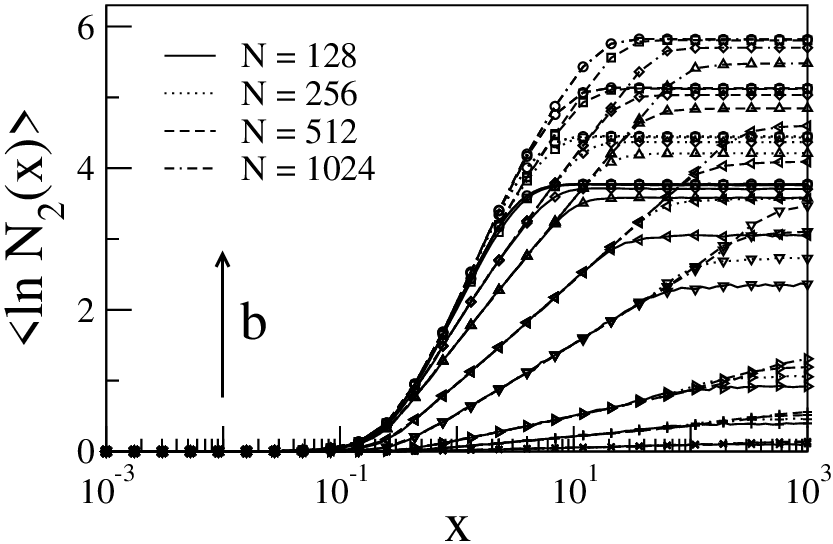}
\caption{$\bra \ln \mathcal{N}_2(x) \ket$ as a function of $x$
for $N=128$, 256, 512, and 1024.
The bandwidth is $b=100 (\circ)$, $40 (\Box)$, $10 (\Diamond)$,
$4 (\bigtriangleup)$, $1 (\lhd)$, $0.4 (\bigtriangledown)$,
$0.1 (\rhd)$, $0.04 ({+})$, and $0.01 ({\times})$.}
\label{fig:fig5}
\end{center}
\end{figure}

In the numerics we assume without loss
of generality that the mean level spacing
is $\Delta=1$. This implies that for $b>1$
the threshold for mixing of levels is
${x_c=\pi^{-1/2}}$, 
while for $b<1$ it is
${x_c=(\pi b)^{-1/2}}$.
Our main interest is in the
non-trivial regime $x > x_c$.
In Fig.~\ref{fig:fig5} we plot $ \ln \overline{\mathcal{N}_2}(x)$
for several values of $b$ and $N$.
We calculate the average using $25\%$ of the
eigenstates at the middle of the spectra
from a number of realizations of $\bm{B}$ summing up
a total of $100,000$ data values for each $(b,N)$.

Looking at Fig.~\ref{fig:fig5} we see that for ${x=0}$
we have ${\ln \overline{\mathcal{N}_2} = 0}$
because all the eigenstates have only one component.
We observe that this zero value persists up
to a point $x=x_0(b)$. In principle one can
argue that $x_0(b)$  should be of the order $x_c$
and (for $b>1$) not larger 
than $x_{\tbox{prt}} \approx \sqrt{b}x_c$.
However, from Eq.~(\ref{Db}) it is clear that
for $b>100$ we already have $D_2 \approx D(\infty) = 1$.
Therefore the $b$ dependence of $x_0$ can be
neglected, and in practice cannot be detected.
We shall see below that for any practical purpose
one can take $x_0\approx 0.15$.

As $x$ is increased beyond $x_0$ the participation
ratio $\mathcal{N}_2$ becomes larger.
As long as $x$ is not too large the $\bm{E}$ term
in the Hamiltonian dominates,
and therefore the size of the matrix is
of no importance. Indeed, we see that
the curves in Fig.~\ref{fig:fig5} are $N$ independent
for small $x$~values.  From this plot we find
that the slope of the curves is given by
\begin{equation}
F(b) \approx \left\{ \matrix{
0  & \mbox{$b<0.1$} \cr
0.57+0.2\ln(b) & \mbox{$b\ge 0.1$}
} \right. \ .
\end{equation}

For large enough $x$, the value
of  $\ln \overline{\mathcal{N}_2}(x)$
saturates to the $x=\infty$ multifractal result.
We call the crossover point $x_{\infty}$.
Using the knowledge of both the $x < x_{\infty}$
behavior (as described in the previous paragraph),
and the $x > x_{\infty}$ behavior (saturation),
we deduce that
\begin{equation}
x_{\infty}(N;b) = x_0 \mbox{e}^{-G(b)/F(b)} N^{D(b)/F(b)} \ .
\end{equation}

\begin{figure}[t]
\begin{center}
    \epsfxsize=8cm
    \leavevmode
    \epsffile{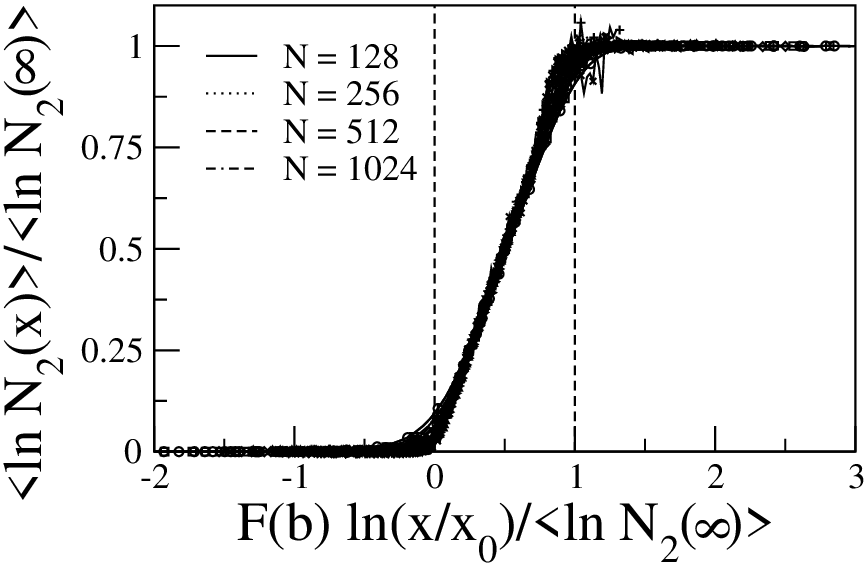}
\caption{The scaled version of Fig.~\ref{fig:fig5}.
$ \bra \ln \mathcal{N}_2(x) \ket / \bra \ln \mathcal{N}_2(\infty) \ket$
is plotted against
$F(b) \ln(x/x_0) / \bra \ln \mathcal{N}_2(\infty) \ket$.
The parametric crossovers at $x=x_0$ and $x=x_{\infty}$,
are indicated by dashed lines.}
\label{fig:fig6}
\end{center}
\end{figure}

Putting together all the above findings
we end up with the following global
scaling relation for the participation ratio:
\begin{equation}
\overline{\mathcal{N}_2}(x) = \left\{
\begin{array}{ll}
\approx 1            & \quad x < x_0 \\
(x/x_0)^{F(b)}   & \quad x_0 < x < x_{\infty}(N;b) \\
\eexp{-G(b)}N^{D(b)}  & \quad x > x_{\infty}(N;b)
\end{array}
\right. .
\label{D2theo}
\end{equation}
In Fig.~\ref{fig:fig6} we demonstrate this scaling. We plot
$\ln \overline{\mathcal{N}}_2(x) / \ln \overline{\mathcal{N}}_2(\infty)$
as a function of the scaled variable
$F(b) \ln(x/x_0) / \ln \overline{\mathcal{N}}_2(\infty)$.
We see clearly the trivial crossover at $x=x_0$
and the non-trivial crossover to multifractality
at $x=x_{\infty}$.

Possibly it is more instructive to describe
the behavior of $\overline{\mathcal{N}_2}$ as a function of $N$
for a given $x$. The interesting case is to have
a fixed value of $x$ which is much larger than $x_0$.
As we increase~$N$ we have a multifractal growth
$\overline{\mathcal{N}_2} \propto N^{D_2}$. This
goes on as long as $x_{\infty}(N;b)$ remains
smaller than $x$. After that $\overline{\mathcal{N}_2}$
reaches saturation, as implied by Eq.~(\ref{D2theo}).
The saturation value is related to the ``width"
of the LDOS, and hence has an algebraic dependence
on the dimensional strength ($x/x_0$) of the perturbation \cite{CH2000}.
It is pleasing to note that for $b\sim1$
we observe $F(b) \approx 2/3$ which is related
to considerations as in \cite{CH2000}.

\section{Conclusions}

We have analyzed the parametric evolution
of eigenstates for the WLRM model. Both, the standard WBRM and the
WLRM models, exhibit a crossover from a perturbative regime
where the LDOS is Lorentzian-like to a non-perturbative
regime where the LDOS is semicircle-like. However, there
is in both cases an additional crossover which is not
captured by the conventional LDOS analysis:
In the case of the standard WBRM model it is the
well studied crossover to an Anderson localization regime,
where the eigenstates become exponentially localized;
In the case of the WLRM model it is the emergence of multifractality.

We have also shown that the WLRM model possess the $x$~invariance
property, absent in the standard WBRM model and in other models
that do not have a Lorenzian band profile. 
Both the non-perturbative crossover and the $x$~invariance
property characterize realistic quantized Hamiltonians.


\ \\

{\bf Acknowledgments.}
This research was supported by a grant from the GIF, the
German-Israeli Foundation for Scientific Research and Development,
and by the Israel Science Foundation (grant No.11/02).


\end{document}